\documentclass[12,manuscript,usenatbib]{aastex}
\usepackage{amsmath}

\newcommand{\kms}{km~s$^{-1}$}
\newcommand{\Msun}{M_{\odot}}
\newcommand{\Lsun}{L_{\odot}}
\newcommand{\kmsMpc}{km~s$^{-1}$~Mpc$^{-1}$}

\shorttitle{Draining Local Void}
\shortauthors{Rizzi et al.}

\begin{document}


\title{Draining the Local Void}


\author{Luca Rizzi}
\affil{W. M. Keck Observatory, Kamuela, HI 96743}
\email{lrizzi@keck.hawaii.edu}

\and

\author{R. Brent Tully}
\affil{Institute for Astronomy, University of Hawaii, Honolulu, HI 96822}

\and

\author{Edward J. Shaya}
\affil{University of Maryland, Astronomy Department, College Park, MD 20743}

\and

\author{Ehsan Kourkchi}
\affil{Institute for Astronomy, University of Hawaii, Honolulu, HI 96822}

\and
\author{Igor D. Karachentsev}
\affil{Special Astrophysical Observatory, Russian Academy of Sciences, N. Arkhz, KChR, 369167, Russia}


\begin{abstract}
Two galaxies that lie deep within the Local Void provide a test of the expectation that voids expand.  The modest ($M_B \sim -14$) HI bearing dwarf galaxies ALFAZOAJ1952+1428 and KK246 have been imaged with Hubble Space Telescope in order to study the stellar populations and determine distances from the luminosities of stars at the tip of the red giant branch.  The mixed age systems have respective distances of 8.39 Mpc and 6.95 Mpc and inferred line-of-sight peculiar velocities of $-114$~\kms\ and $-66$~\kms\ toward us and away from the void center.  These motions compound on the Milky Way motion of $\sim 230$~\kms\ away from the void.  The orbits of the two galaxies are reasonably constrained by a numerical action model encompassing an extensive region that embraces the Local Void.  It is unambiguously confirmed that these two void galaxies are moving away from the void center at several hundred \kms.
\end{abstract}





\section{Introduction}

According to theory voids expand, but what is the observational support?  The most convincing evidence until now has been provided by our own motion with respect to the Local Void.  Our Galaxy and most of our neighbors lay in a flattened structure called the Local Sheet \citep{2008ApJ...676..184T} which is a wall of the Local Void \citep{1987nga..book.....T}.  With the discovery of the cosmic microwave background dipole \citep{1969Natur.222..971C, 1971Natur.231..516H} that infers a Local Group motion of 631~\kms\ \citep{1996ApJ...473..576F} there was early speculation that a component might come from motion away from the Local Void \citep{1988lsmu.book..199L}.  There have been hints of Local Void expansion at the level of $\sim 300$~\kms\ from the distances to galaxies at the periphery of the void \citep{2011Ap.....54....1N, 2014Ap.....57..457K}.

This speculation has been strongly confirmed.  Accurate distances to nearby galaxies can be acquired with the Tip of the Red Giant Branch (TRGB) method from imaging with Hubble Space Telescope \citep{2009AJ....138..332J}.  Distances with accuracy $\sim5\%$ can be obtained for galaxies within 10~Mpc and, to date, there are measures for almost 400 galaxies.\footnote{http://edd.ifa.hawaii.edu/ see catalog CMDs/TRGB}  There is a clear convergence of the Local Sheet with filaments in the opposite direction from the Local Void, those \citet{1987nga..book.....T} called the Leo Spur, and the Antlia and Doradus clouds.  These structures are approaching each other at $\sim 260$~\kms\ \citep{2008ApJ...676..184T, 2015ApJ...805..144K}.  Viewed in reference frames averaged over 20 Mpc or greater, it is evident that most of this convergence is due to a downward motion of the Local Sheet, away from the Local Void.  See Figures $6-10$ in \citet{2008ApJ...676..184T}.

While it is compelling that the Local Sheet is displacing away from the Local Void, it can be expected that the dynamics are complicated.  There is a void on the opposite side of the Local Sheet, though smaller.  If the opposing voids were equal there should be no displacement normal to the sheet.  The problem has been studied by \citet{2009LNP...665..291V} with simulations and extensively reviewed by \citet{2011IJMPS...1...41V}.  They note that the place to look for uncompromised expansion is not in the walls but in the voids themselves.

The Local Void is the only viable place to look for direct evidence of void expansion because a target must be at a small distance, $d$, to unambiguously disentangle the cosmic expansion, $H_0 d$, and a peculiar component, $V_{pec}$, of observed velocities, $V_{obs}$:
\begin{equation}
V_{pec} = V_{obs} - H_0 d  .
\label{Eq:vpec}
\end{equation}
Unfortunately there are not many galaxies in voids.  Until now, only one decent case in the Local Void has been studied, KK246 = ESO461-036 = PGC64054 \citep{1998A&AS..127..409K, 2006AJ....131.1361K, 2011AJ....141..204K}.  It was noted that KK246 has a substantial apparent motion of expansion away from the Local Void \citep{2008ApJ...676..184T}.  However, the argument from that case alone was not solid because (a) the angle between the direction toward KK246 and the void center is sufficiently large ($55^{\circ}$) that only a modest fraction of radial expansion from the void projects toward us, and (b) the distance to KK246 had a relatively large uncertainty due to reddening.

The difficulty in finding alternative targets is not just the paucity of galaxies in voids.  By happenstance, the center of the Local Void is behind the Galactic center.  Still, HI surveys confirm that there are few galaxies in the region of the Local Void \citep{2016AJ....151...52S}.  With considerable fortune, though, a small, low redshift galaxy ($V_{helio} = 279$~\kms) was revealed by the blind HI survey with Arecibo Telescope \citep{2011ApJ...739L..26M} that lies in a relatively clear window and is nicely aligned ($10^{\circ}$) toward the center of the Local Void.  This galaxy, ALFAZOAJ1952+1428 (Principal Galaxies Catalog: PGC5060431), is the main subject of the ensuing discussion.

\section{Observations}


Two galaxies have been observed in the course of HST program GO-14167 (PI: E. Shaya).  ALFAZOA1952+1428 was observed with Advanced Camera for Surveys (ACS) for 4508s and 4640s in F814W and F606W, respectively over 4 orbits.  In addition KK246, which had already been observed with ACS in these same filters in one orbit (GO-9771; PI: I. Karachentsev) was now observed in one further orbit with the infrared side of Wide-Field Camera 3 (WFC3/IR), with 1306s exposures in each of F160W and F110W.  General parameters describing the two targets are provided in Table 1.

\begin{table}
\begin{center}
\caption{Basic Parameters}
\bigskip
\begin{tabular}{lcc}
\hline
PGC           & 5060431     & 64054   	  \\
Name          & ALFAZOA     & KK246   	  \\
RA     [d am as]& 19 52 11.8  & 20 03 57.4  \\
Dec    [d m s]  & +14 28 24   & $-$31 40 54 \\
Glon   [deg]  & 52.8241     & 9.7286  	  \\
Glat   [deg]  & $-$6.4205   & $-$28.3689  \\
SGL    [deg]  & 293.2312    & 225.6110 	  \\
SGB    [deg]  & 76.1456     & 39.8077 	  \\
$E(B-V)$ [mag]  & 0.28        & 0.44    	  \\
$B^b$    [mag]  & 15.9        & 15.2    	  \\
$(m-M)_{opt}$ [mag]  & $29.62\pm0.14$& $29.18\pm0.11$\\
$(m-M)_{IR}$ [mag]  &             & $29.24\pm0.12$\\
Dist   [Mpc]  & $8.39\pm0.56$ & $6.95\pm0.29$ \\
$M_B^b$  [mag]  & $-$13.7     & $-$14.0 	  \\
$V_{helio}$ [\kms] & 278         & 427     	  \\
$V_{LS}$    [\kms] & 515         & 455     	  \\
$V_{pec}$   [\kms] & $-114\pm42$ & $-66\pm22$  \\
\hline
\end{tabular}
\end{center}
\label{table1}
\end{table}




\section{Tip of the Red Giant Branch Distances}

\citet{1990AJ....100..162D} and \citet{1993ApJ...417..553L} were early proponents of the standard candle nature of the Helium flash that caps the evolution of stars up the Red Giant Branch.  The latter authors, with refinements by \citet{1996ApJ...461..713S}, used the Sobel filter to identify the luminosity of the Tip of the Red Giant Branch.  \citet{2002AJ....124..213M} developed a maximum likelihood procedure that assumes a step in luminosity at the TRGB and an RGB luminosity function with slope in magnitudes of 0.3 faintward of the tip.  \citet{2009ApJ...690..389M} have emphasized the requirements for access to an adequate number of stars near the tip for a robust measurement.

Our own procedure follows the maximum likelihood method of \citet{2006AJ....132.2729M}.  Artificial stars with a wide range of magnitudes and colors are superimposed on images in order to evaluate recovery rates and uncertainties at representative levels of crowding.  The stellar photometry is carried out with the DOLPHOT package (\citet{2000PASP..112.1383D} and updates).  The returns from this stage of the analysis are color-magnitude diagrams (CMD) and TRGB magnitudes and colors.  With the targets under discussion the foreground extinctions from our Galaxy are significant and reddening is estimated from the color displacements of the zero age main sequence in the CMD.

The color of the RGB is strongly metallicity dependent and, to a lesser degree, age dependent.  Spectral energy shifts to the infrared with both increasing metallicity and age: whence flux at blueward optical bands drop and flux at infrared bands increase sharply.  Observations at $I$ band, or the nearly equivalent $F814W$ band with HST, are attractive because the TRGB magnitude trends decrease only slightly with increasing metallicity or age.  In the HST infrared bands of interest, $F110W$ and especially $F160W$, the TRGB magnitudes increase strongly with increasing metallicity or age.  The sweet spot where metallicity and age effects are minimized lies around $0.9-1$~micron.  The $F814W$ filter is a good approximation.  In the infrared, by $F160W$ close attention must be given to the color term but, in compensation, the TRGB is very bright!  The calibrations used in the present paper for our imaging at $F814W$ and $F606W$ with ACS follows the procedures given by \citet{2007ApJ...661..815R}, including metallicity and reddening corrections and translations to $I$ and $V$ colors.  For the calibration of imaging at $F160W$ and $F110W$ with WFC3/IR we follow \citet{2014AJ....148....7W}.

\section{ALFAZOAJ1952+1428}

\subsection{Distance}

\begin{figure}
\epsscale{1.2}
\plotone{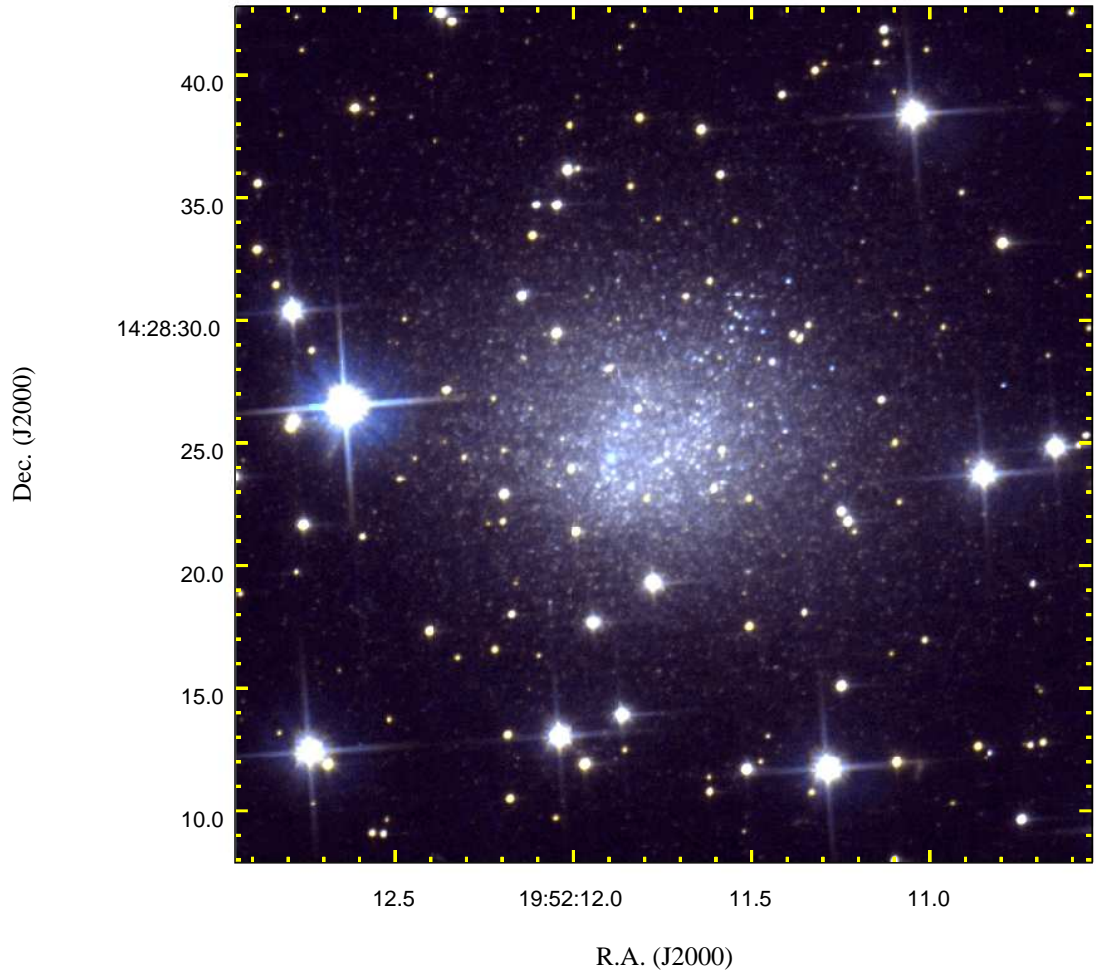}
\caption{Color image of ALFAZOAJ1952+1428 built from HST F606W and F814W exposures.}
\label{color}
\end{figure}

\begin{figure}
\epsscale{.80}
\plotone{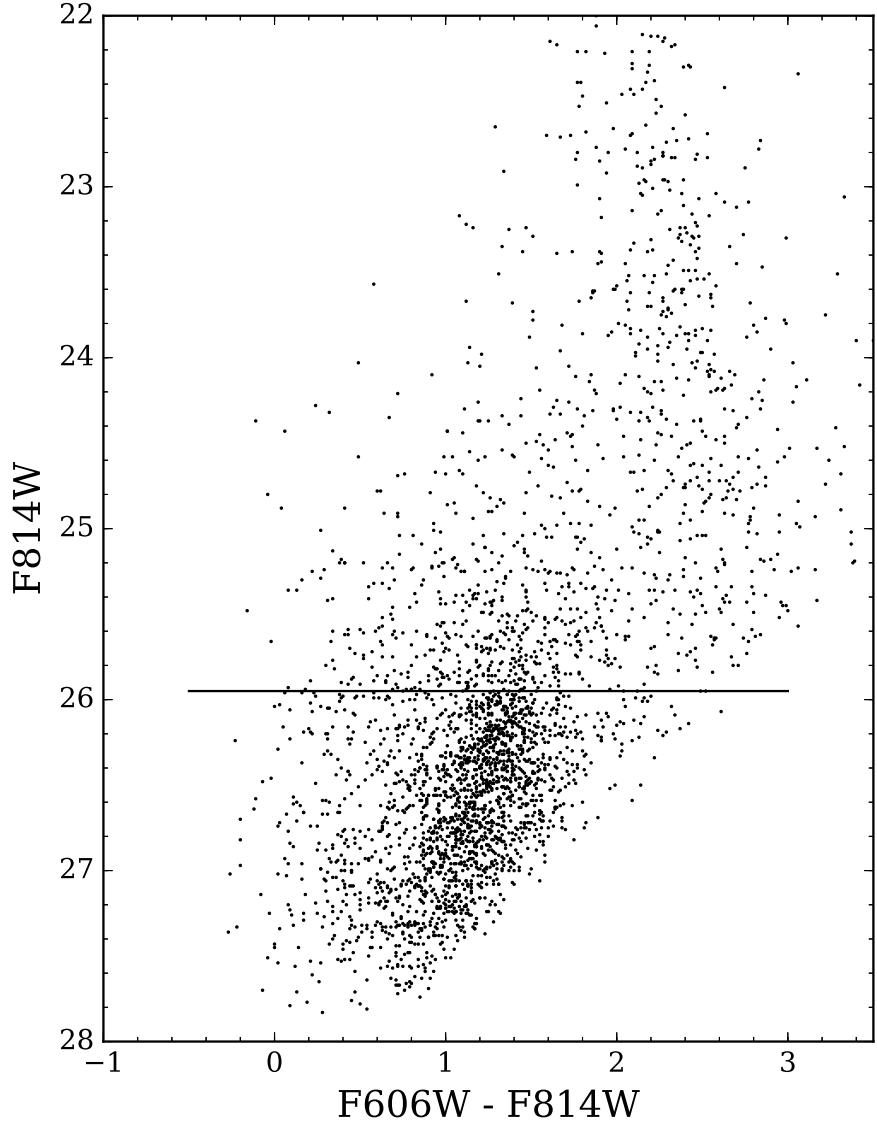}
\caption{Color magnitude diagram for ALFAZOAJ1952+1428.  The TRGB is located at $F814W = 25.95$ as indicated by the horizontal line.  The points redward of $F606W-F814W = 1.5$ and brighter than $F814W = 25$ represent foreground stars, numerous at the low latitude of the target.}
\label{cmd1}
\end{figure}

\begin{figure}
\epsscale{.80}
\plotone{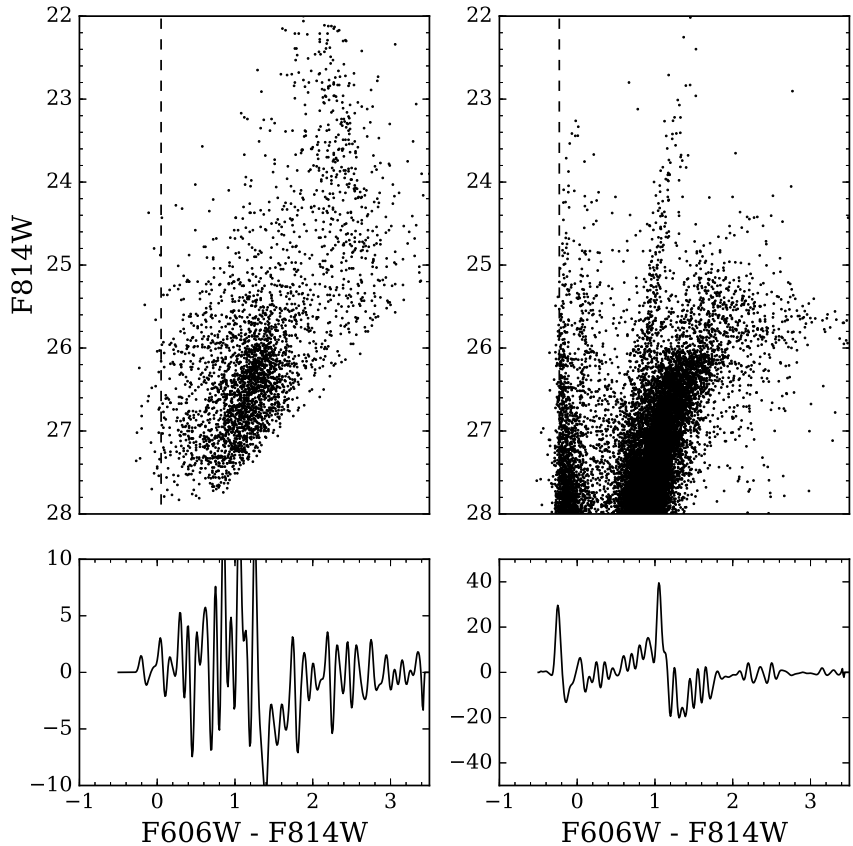}
\caption{Reddening difference between ALFAZOAJ1952+1428 (left) and NGC~300 (right).  The NGC~300 CMD has been shifted in magnitude so the levels of the TRGB agree.  The results of a Sobel filter analysis are demonstrated in the lower panels. The first significant peaks to the left in the Sobel filter responses identify the colors of the zero age main sequence in the two CMD.  The vertical dotted lines lie at these colors.  The displacement in color of this feature between the two CMD is a measure of the relative reddening between the two targets.  The reddening in the case of NGC~300 is very small.} 
\label{reddening1}
\end{figure}

Results are summarized in a series of figures.   Figure~1 is a color image of ALFAZOAJ1952+1428 (for a high resolution image, see the entry PGC5060431 in the CMDs/TRGB catalog in the Extragalactic Distance Database at http://edd.ifa.hawaii.edu).  The CMD built with components of this image is shown in Figure~\ref{cmd1}.  The TRGB is determined to lie at $F814W = 25.95\pm0.09$ as indicated by the horizontal line.  The tip color is $F606W - F814W = 1.32\pm0.08$.  These quantities are affected by reddening.  The amplitude of extinction is estimated by the color displacement of the zero-age main sequence.  Figure~\ref{reddening1} demonstrates the procedure in a comparison with a CMD obtained from observing NGC~300 in the same filters.  The color displacement is $(A_{606}-A_{814})^{ALFAZOA} - (A_{606}-A_{814})^{N300} = 0.28\pm0.04$ (The color displacement in the case of NGC~300 at the south polar cap is a very small 0.008 mag).     Assuming the \citet{1999PASP..111...63F} reddening relations with $R_V=3.1$, then $E(B-V)^{ALFAZOA} = 0.28\pm0.04$ and $A_{814}=0.43\pm0.10$ (compared with the  \citet{2011ApJ...737..103S} value of  0.40), so the reddening adjusted distance modulus is $(m-M)_0 = 29.62\pm0.14$ corresponding to the distance $8.39\pm0.56$~Mpc.

The HI heliocentric velocity $V_h = 278\pm1$~\kms\ transforms to the Local Sheet frame \citep{2008ApJ...676..184T} as $V_{LS} = 515$~\kms.  The Local Sheet frame is a variant of the Local Group frame of which there are three popular renditions \citep{1977ApJ...217..903Y, 1996AJ....111..794K, 1999AJ....118..337C}.  Taking an average, $V_{LG} = 492$~\kms.  If $H_0 = 75$~\kmsMpc\ (Tully et al. 2016) then $H_o d = 629\pm 42$~\kms.  Then from Eq.~\ref{Eq:vpec} $V_{pec} = -114\pm42$~\kms\ in the Local Sheet frame or $-137$~\kms\ in the averaged Local Group frame.  A choice of a different value for the Hubble Constant of $\Delta H_0 = -1$ changes $V_{pec}$ by only $+7$~\kms.  The implications of this measurement of $V_{pec}$ will be considered in the discussion section.

\subsection{Properties}

\begin{figure}
\epsscale{.80}
\plotone{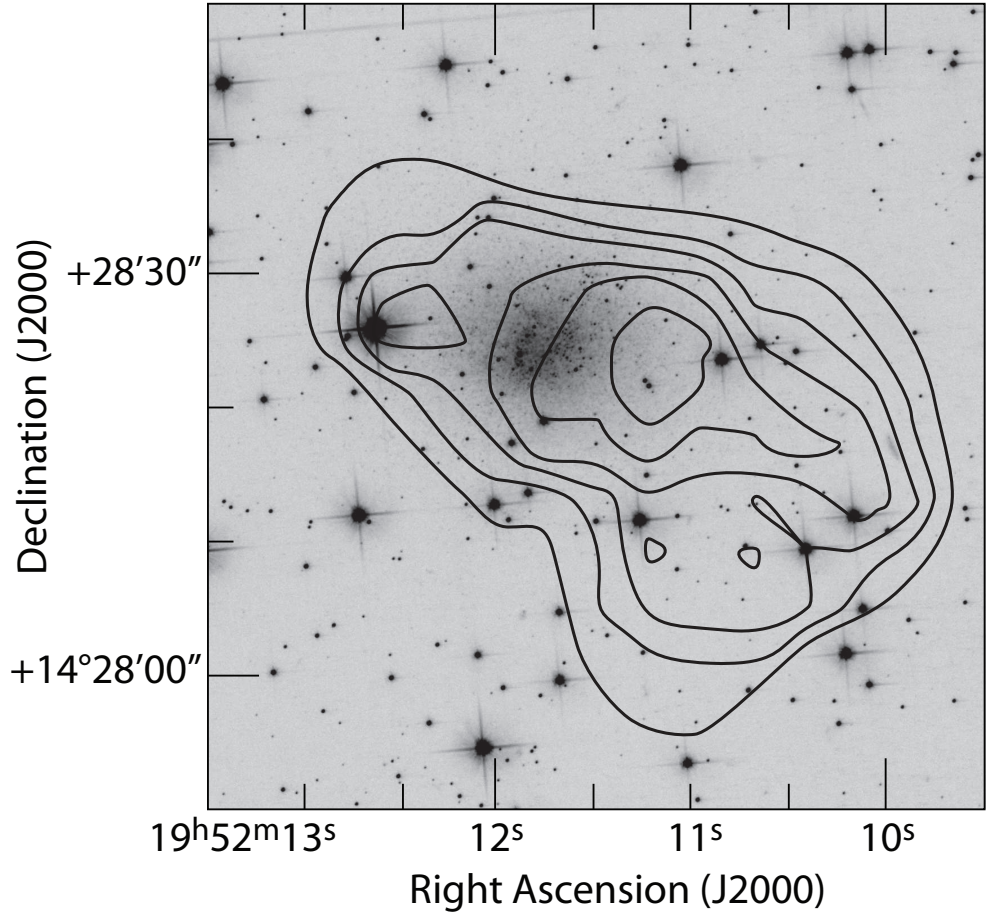}
\caption{Image of ALFAZOAJ1952+1428 with superimposed HI contours.}
\label{HI}
\end{figure}

\begin{figure}
\epsscale{1}
\plottwo{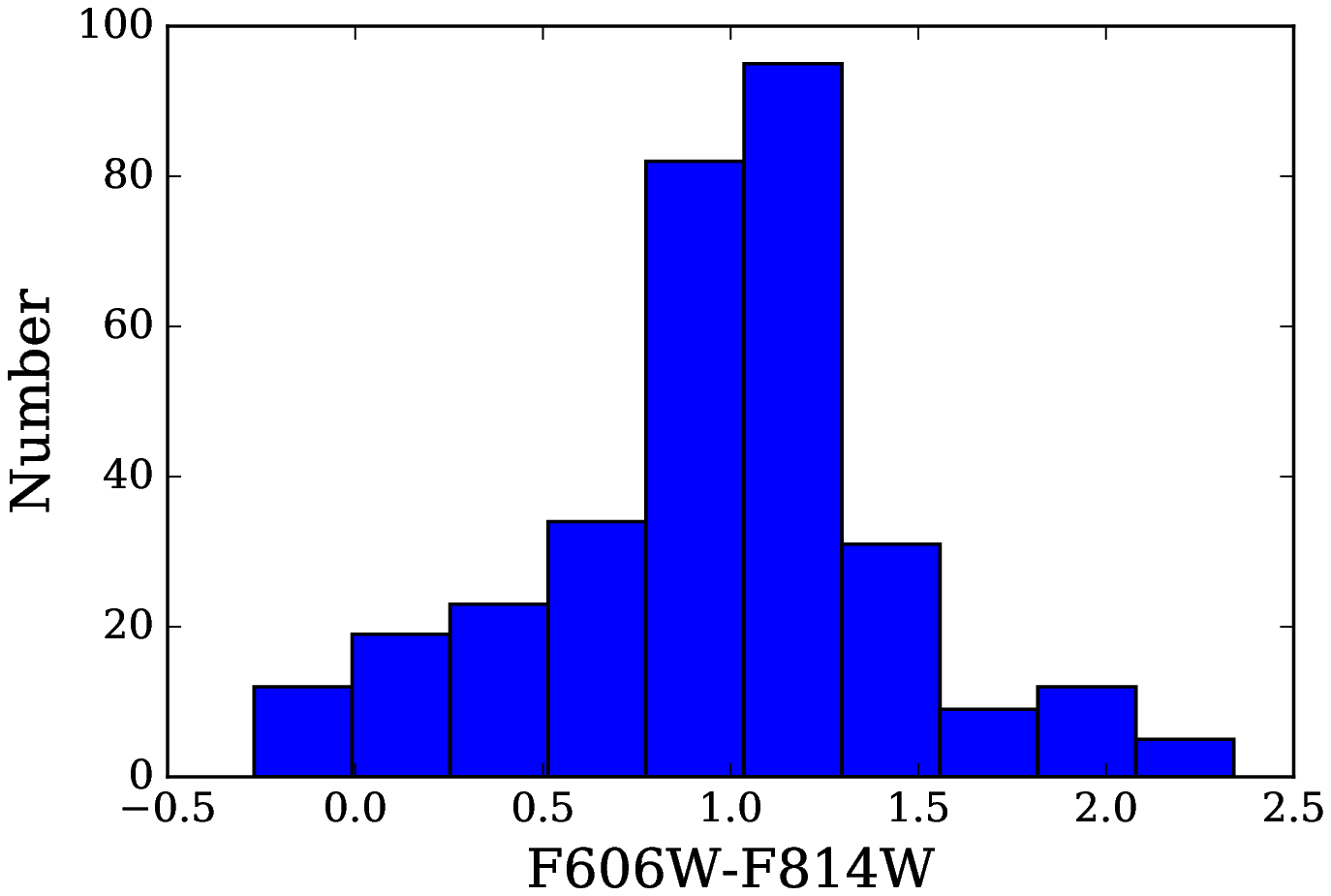}{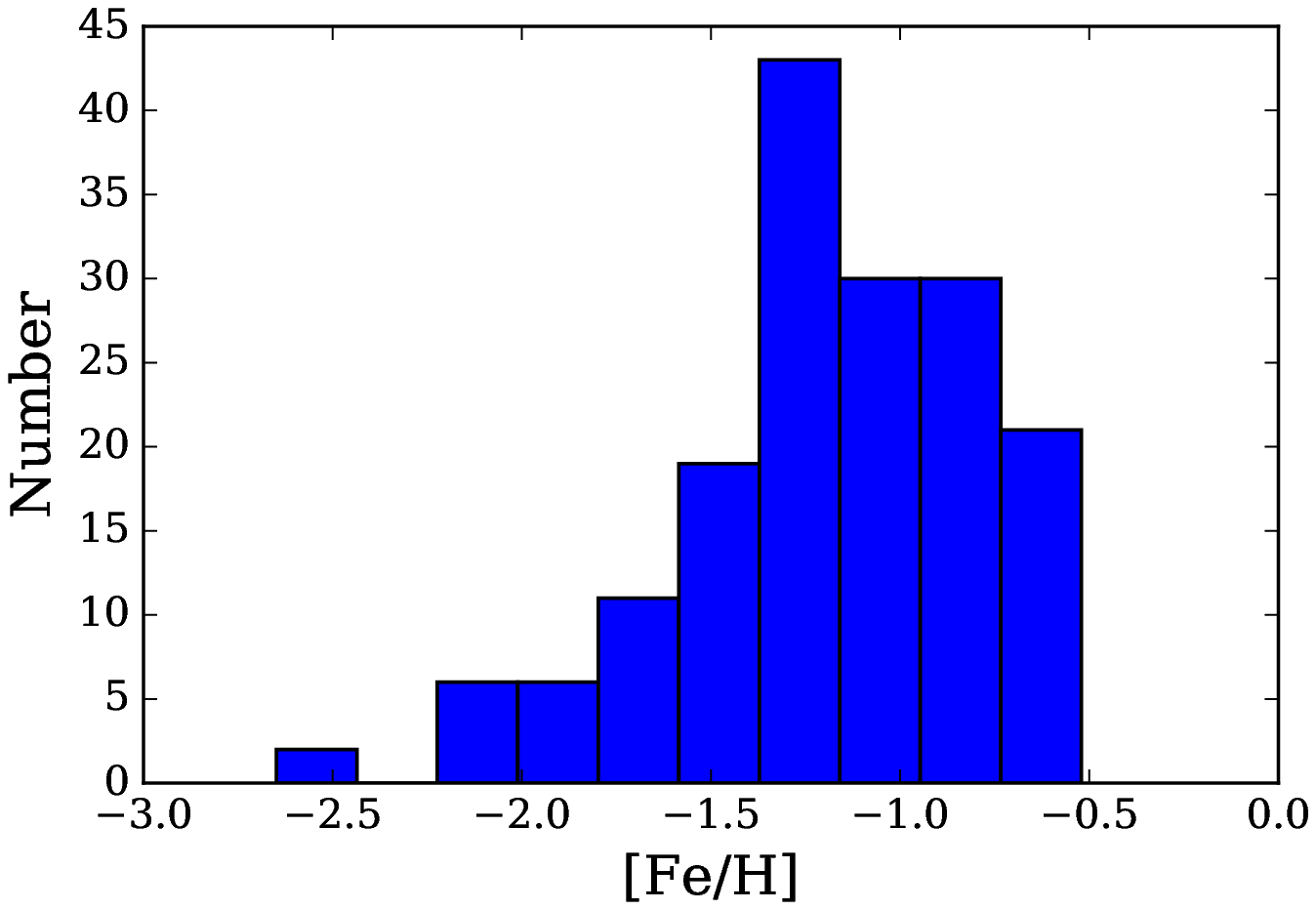}
\caption{ALFAZOAJ1952+1428 RGB reddening corrected colors (left) and implied metallicities (right).} 
\label{metal}
\end{figure}

A neutral Hydrogen detection at a small redshift drew attention to ALFAZOAJ1952+1428.  The galaxy has been observed with the Extended Very Large Array (EVLA) by \citet{2011ApJ...739L..26M} and contours of their HI map are superimposed on the HST image of the galaxy in Figure~\ref{HI}.  The HI emission is extended over a diameter of $50^{\prime\prime} \sim 2$~kpc, comparable to the 2~kpc$\times$1.2~kpc extent of the optically detected galaxy.  The peak of the HI emission is seen to be displaced from the peak of the stellar concentration by 330~pc.  It is evident from the galaxy image in Fig.~\ref{color} that there are a large number of relatively young stars concentrated within a region of 250~pc near the optical center.  The most luminous blue stars in the CMD have ages $\sim 100$~Myr.  Intermediate age (few Gyr) Asymptotic Giant Branch stars are plentiful.  The RGB contains stars greater than 1~Gyr old.  The CMD is too shallow to inform us about whether there is or is not an ancient population.  The RGB is wide, indicating that star formation has been extended or has occurred in multiple episodes.  The mean reddening corrected color of the TRGB of $F606W-F814W = 1.02$ is quite blue implying a low metallicity.  The left panel of Figure~\ref{metal} is a histogram is the reddening corrected colors of RGB stars from the TRGB magnitude to 0.2 mag faintward of the RGB tip.  The histogram in the right panel is a translation from color to metallicity based on Padova isochrones \citep{2012MNRAS.427..127B} and an age of 12 Gyr.  The median of the distribution is [Fe/H] $= -1.16$, greater than an order of magnitude more metal poor than solar. 

\citet{2013AstBu..68..381K} imaged ALFAZOAJ1952+1428 in the H$_{\alpha}$ line and interpreted the flux as the star formation rate (SFR) and mass normalized specific star formation rate (sSFR).  Given the distance 8.39~Mpc and reddening $E(B-V)=0.28$, then ${\rm log}(SFR) = -2.94~\Msun {\rm yr}^{-1}$ and ${\rm log}(sSFR) = -10.72~{\rm yr}^{-1}$.  This level of star formation is moderate.

\begin{figure}
\plotfiddle{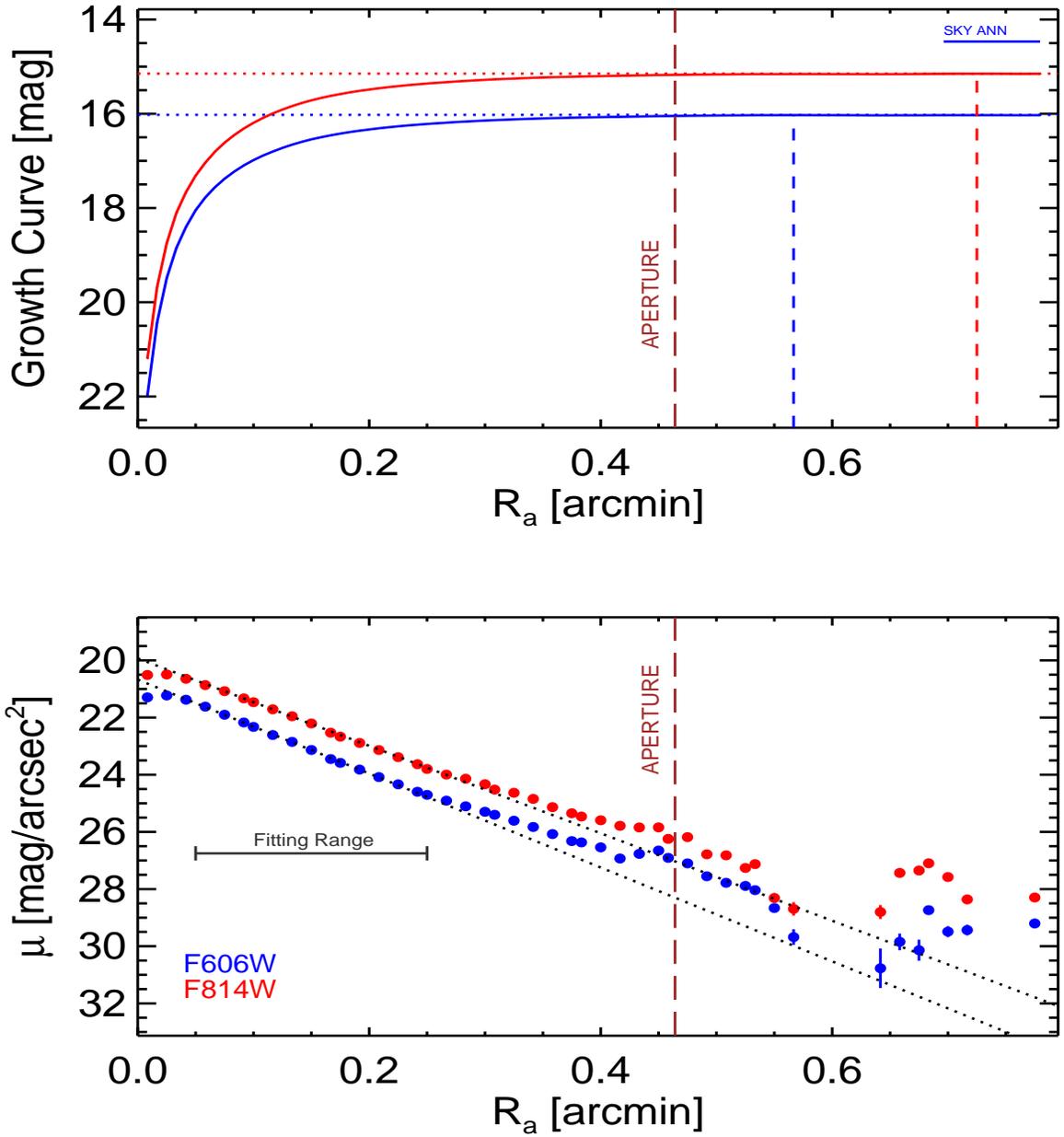}{2cm}{90.}{500}{500}{0}{25}
\caption{F606W and F814W magnitudes (top) and surface brightness (bottom) as a function of radius.}
\label{photom}
\end{figure}

Globally, the apparent magnitude reported by \citet{2011ApJ...739L..26M} after correction for our reddening is $B=15.9$, $M_B^b = -13.7$.  We find asymptotic magnitudes $V = 16.20\pm0.10$ and $I = 15.14\pm0.05$ from the HST aperture photometry growth curves shown in the top panel of Figure~\ref{photom}.  The conversions from HST flight magnitudes follow the prescriptions given by \citet{2005PASP..117.1049S}.  It is seen in the bottom panel that there is a classic exponential radial dependence of the surface brightness peaking at the center at 20.67 in the F606W band ($\mu_0(V) = 20.80\pm0.09$) and 19.93 in the F814W band ($\mu_0(I) = 19.92\pm0.05$).  The exponential scale length averaged across the two bands is 4.1 arcsec = 170 pc.  Dereddened, $I^b = 15.57$ and $M_I^b = -14.05$.  The galaxy is a modest dwarf, compact enough to be considered a BCD, Blue Compact Dwarf, although from the ages of the youngest stars it is post-starburst.  \citet{2011ApJ...739L..26M} point out that the HI to luminosity ratio $M_{HI}/L_B = 0.3~\Msun/\Lsun$ is modest.

ALFAZOAJ1952+1428 is of particular interest because it lies in an extremely low density neighborhood.  The properties that have been identified are not unusual for a gas-bearing dwarf galaxy.

\section{KK246}

A TRGB measurement of the distance to KK246 and follow-up observations have been reported by \citet{2006AJ....131.1361K} and \citet{2011AJ....141..204K}.  We had considered the distance to be somewhat uncertain because, due to significant foreground obscuration, the TRGB at $F814W$ lies only $\sim 0.8$ mag above the photometric limit set by $S/N \sim 5$ with the single orbit HST observation that was available.  \citet{1995AJ....109.1645M} have pointed out concerns with TRGB measurements when less than 1 magnitude at the top of the RGB is accessible.
We appreciated that, because of reduced extinction in the infrared, the TRGB would be comfortably detected with a one orbit HST observation split between $F110W$ and $F160W$ with WFC3/IR.  That expectation was confirmed with the new observations that will be discussed.

\begin{figure}
\epsscale{1.2}
\plottwo{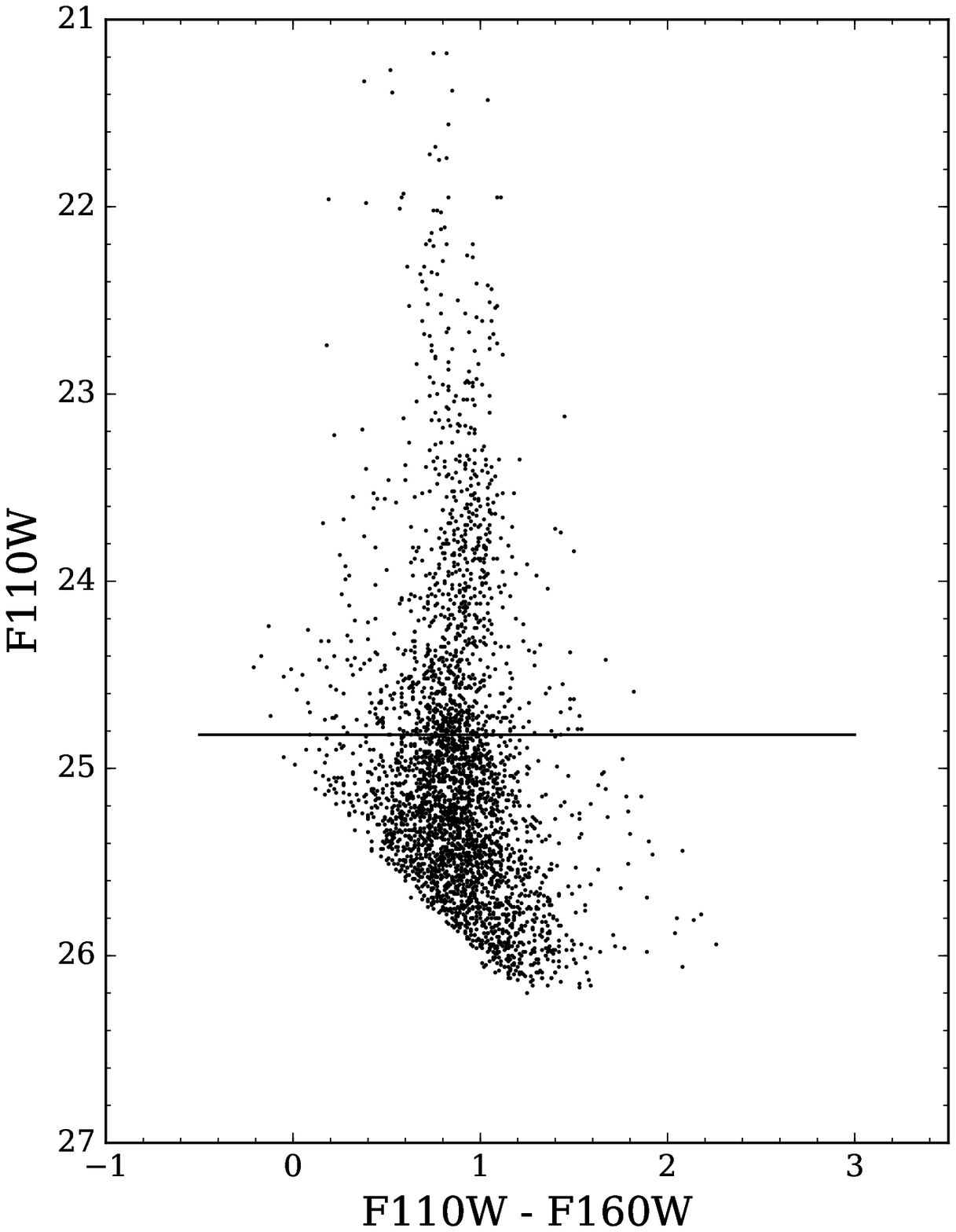} {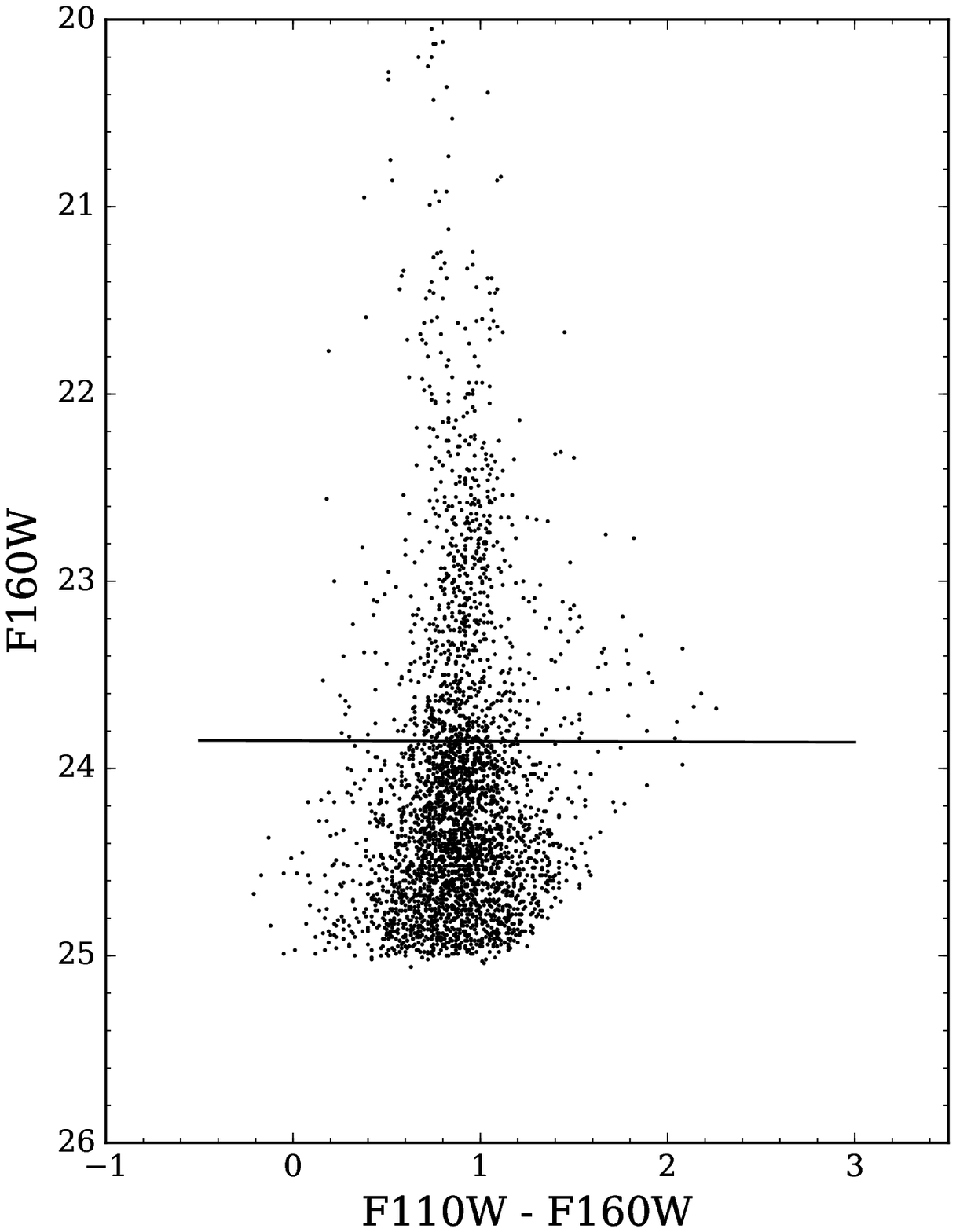}
\caption{Alternate color magnitude diagrams with $F110W$ and $F160W$ magnitudes.  In each case the TRGB level is identified by a horizontal line.}
\label{cmd2}
\end{figure}

\citet{2014AJ....148....7W} provide TRGB calibrations for both the $F110W$ and $F160W$ bands.  CMDs with each band in the ordinate are shown in the two panels of Figure~\ref{cmd2}.  Of course, the same stars contribute to both variations.  It is worth analyzing the two CMD separately, though, because the effects of the color terms and reddening are different.  $F110W$ is more affected by reddening but metallicity effects are smaller.  $F160W$ has the reverse advantage and disadvantage.

As anticipated, the TRGB is slightly more than 1 mag above the faint threshold in both CMD variants.  We find $F110W_{TRGB} = 24.82 \pm0.06$ and $F160W_{TRGB} = 23.86 \pm0.04$.  These values are affected by extinction.  Stars are seen blueward of the RGB in Fig.~\ref{cmd2} and could give a limit on the location of the zero age main sequence but the statistics are poor.  A reddening of $E(B-V) =0.44 \pm0.10$ is better determined by the main sequence identification in the pre-existing ACS $F814W$ and $F606W$ observations.  A high resolution image and optical CMD of KK246 can be seen at http://edd.ifa.hawaii.edu by selecting the CMDs/TRGB catalog and searching for KK246=PGC64054.  The \citet{2011ApJ...737..103S} reddening value of $E(B-V) = 0.26$ is almost certainly too small in this instance.

The de-reddened TRGB values at $F110W$ and $F160W$ respectively are $24.43\pm0.11$ and $23.63\pm0.06$.  The de-reddened tip color is $F110W_0 - F160W_0 = 0.80\pm0.09$.  From   \citet{2014AJ....148....7W} the absolute tip magnitudes are $-4.81\pm0.06$ and $-5.61\pm0.06$.  The resultant distance moduli are $29.24\pm0.12$ and $29.24\pm0.09$.  The values from the two bands agree and give the distance $7.05\pm0.40$~Mpc.

The new distance measurement is in agreement with the earlier optical measurement of $(m-M)_{814} = 29.18\pm0.11$, $d_{814} = 6.86$~Mpc.  Averaging the optical and infrared values, $(m-M)_0 = 29.21\pm0.09$, $d = 6.95\pm0.29$.  The product $H_0 d = 521\pm22$~\kms\ exceeds the observed $V_{LS} = 455$~\kms\ ($V_h = 427$~\kms), so the peculiar velocity is negative at $V_{pec} = -66\pm22$~\kms.  The averaged local group rest frame velocity is 455~\kms\ so $V_{pec} = -76$~\kms\ in that frame.  $\Delta H_0 = -1$ causes $\Delta V_{pec} = +6$~\kms.

KK246 has a far ultraviolet flux from the space satellite GALEX.  Following \citet{2013AstBu..68..381K} and accepting a distance 6.95~Mpc and $E(B-V) = 0.44$ then ${\rm log}(SFR) = -2.42~\Msun {\rm yr}^{-1}$ and ${\rm log}(sSFR) = -10.25~{\rm yr}^{-1}$.  Both galaxies observed in the Local Void have moderate star formation rates.   

\citet{2011AJ....141..204K} report observations of KK246 with the VLA.  Neutral Hydrogen is much more abundant and extensive than in the case of ALFAZOAJ1952+1428.  The HI extends over 6~kpc, 5 times farther than the optical galaxy and there is a well developed rotation pattern.  With our revision of the reddening and distance for KK246, the galaxy can be described by the parameters $M_B^b = -14.0$, $M_{HI} = 8.3 \times 10^7~\Msun$, $M_{HI}/L_B = 1.3~\Msun/\Lsun$, and $M_{dyn}/L_B = 57~\Msun\Lsun$.  This latter figure, the dynamical mass from the rotation curve to blue light ratio, is reduced from the value given by Kreckel et al. but it remains that KK246 is very dark.

\citet{2013AJ....145..101K} assign a `tidal index' to nearby galaxies as a measure of their individual isolation.\footnote{Updated at http://www.sao.ru/lv/lvgdb}  The two galaxies of this study have the lowest tidal index parameters of any galaxies in the volume within 10~Mpc. 

\section{Discussion}

ALFAZOAJ1952+1428 and KK246 are the only two galaxies known to exist in the Local Void that have been suspected to lie within 10 Mpc, the domain fruitfully accessed with HST imaging.  Now, such observations provide CMD that place them at 8.39 and 6.95 Mpc respectively.  The two galaxies have similar integrated magnitudes of $M_B^b \sim -14$ and both have experienced star formation over the last few hundred million years.  KK246 is enveloped in a much more extensive HI cloud than ALFAZOAJ1952+1428 and the regular rotation of that cloud gives evidence for the existence of an extended dark halo.  The HI associated with ALFAZOAJ1952+1428 is too restricted to provide insight regarding a halo.  In other respects, the two galaxies resemble other dwarfs containing HI. 

The particular interest in this study is in the peculiar motions of the two galaxies.    After cancelation of the cosmic expansion, both galaxies have apparent motions toward us.  Our hypothesis is that we are seeing projections of motions away from the center of the Local Void, the superHubble expansion expected in void regions where outflow velocities are proportional to the distance from the void center \citep{2011IJMPS...1...41V}.  In the case of ALFZOAJ1952+1428, almost the full amplitude of the peculiar velocity of $-114$~\kms\ would be aligned with the void center.   KK246, on the other hand, lies at an angle of $\sim 55^{\circ}$ from the direction of the center of the Local Void so the observed peculiar motion of $-66$~\kms\ might reflect a motion of $\sim -115$~\kms\ directed radially away from the void.  The uncertainties with these measures are $\sim 40$~\kms.

These peculiar velocities are with respect to us but we are moving, evidently participating in the void expansion.  It was stated in the introduction that the Local Sheet is displacing toward structures opposite the Local Void at $\sim 260$~\kms.  It follows that at least ALFAZOAJ1952+1428 has a deviant motion of $\sim 375$~\kms\ toward those features.   Without linkage to a larger scale reference frame it could be entertained that a component of this velocity is due to motion of the structures below the Local Sheet toward us.  

Such a linkage with large scales is possible.  The {\it Cosmicflows-2} and -3 compendiums supply distances, hence peculiar velocities through Eq.~\ref{Eq:vpec}, for 8,000 and 18,000 galaxies respectively within $z=0.1$ \citep{2008ApJ...676..184T, 2016AJ....152...50T}.  The velocities have been interpreted as 3D maps of densities using the Wiener Filter technique with constrained realizations \citep{2012ApJ...744...43C, 2014Natur.513...71T}.  Then in greater detail locally where non-linear dynamics are important, galaxy orbits have been reconstructed with NAM, numerical action methods \citep{1989ApJ...344L..53P, 1995ApJ...454...15S, 2000MNRAS.313..587N, 2013MNRAS.436.2096S}.  Specifically, orbits are followed for 1385 entities within 38 Mpc, embedded in the Wiener Filter map that has an empty cavity within 38 Mpc.\footnote{As a matter of detail, the Wiener Filter map was derived from the {\it Cosmicflows-2} data while the non-linear NAM analysis makes use of the {\it Cosmicflows-3} data.  The Wiener Filter map provides a tidal field acting on the inner core that is important.}
There is no attempt to resolve the complex orbits within collapsed structures $-$ the `entities' in the model are the halos of groups or individual galaxies outside of groups.  Further, for the entities to be included, they must be either massive enough to be dynamically important or have known distances with accuracies that constrain the model fit.  The importance of the Wiener Filter model of large scale influences is to be appreciated, because the emptiness of the Local Void requires description beyond the 38 Mpc inner domain and because external tides are important \citep{2005A&A...440..425R}.
Details of the study are described by Shaya et al. (in preparation).

The reconstructed orbits are physically plausible but they are not unique.  However there are robust aspects of the best models.  Masses of the major components are reasonable well constrained and the overall flow patterns are preserved.  In particular, nearby where the data density is high the flow patterns are clear.

\begin{figure}
\epsscale{0.7}
\plotone{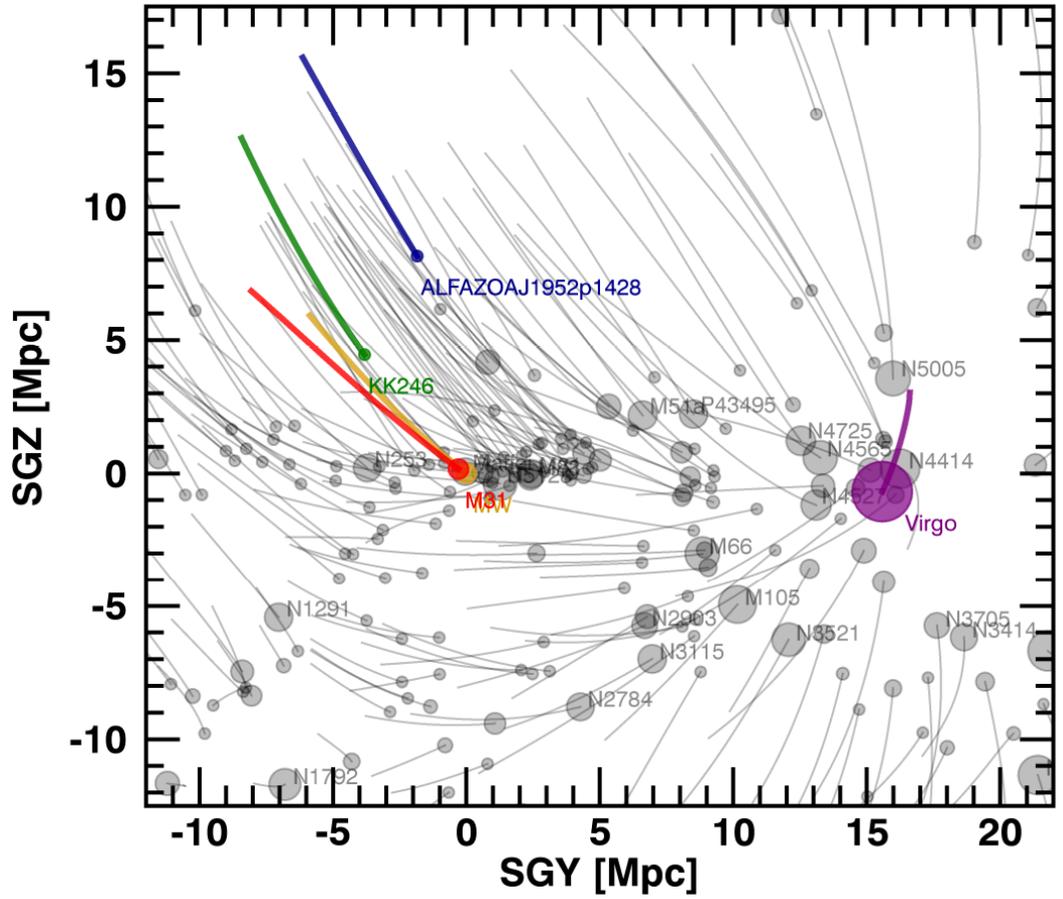} 
\caption{SGY-SGZ ($-6<{\rm SGX}<6$~Mpc) projection of orbits from a numerical action model in co-moving coordinates.  Individual components of the model are located today at the positions of the spheres and the tails indicate orbits since $z=4$.  Elements of particular interest are given colors.}
\label{yz}
\end{figure}

Figure~\ref{yz} illustrates the essential elements of the best orbit models.  The scene is in supergalactic coordinates and illustrates a slab cut out of the full reconstruction.  The Milky Way is at the origin.  The important morphological elements are the Local Sheet \citep{2008ApJ...676..184T} occupying the supergalactic equator with the Virgo Cluster in this dominant plane at the extreme right, and the disjoint Leo Spur and Antlia-Dorado clouds\footnote{The names of structures below the equatorial plane were given by \citet{1987nga..book.....T}.} lying below the equator (negative SGZ).  The domain of the Local Void is above the equatorial band (positive SGZ).  The two galaxies of interest for the current study are located in this region.

The orbits of the specific model are shown in the frame of the center of mass of the ensemble and in co-moving coordinates.  It is evident that there is considerable coherence.  Entities in our vicinity are all plunging down, away from the Local Void, and swinging right, toward the Virgo Cluster.  Motions toward the Virgo Cluster increase as the cluster is approached.  The dominate motion of entities in the Leo Spur (lying at positive SGY, negative SGZ) is toward the Virgo Cluster, a positive SGY motion with a slight upward SGZ trend.  The Antlia-Dorado component (at mostly negative SGY, negative SGZ) is trending slightly downward (toward the Fornax Cluster, off the map to the lower left).  Our two targets of interest in the Local Void have substantial downward motions in the model, following the paths of the Local Sheet systems.  

In detail, our NAM models slightly under predict the motions of the void galaxies toward $-$SGZ.  It was found above that both void galaxies have negative peculiar velocities with respect to the Local Group with roughly $3\sigma$ significance.  However there is the complexity that the galaxies of the Local Group are moving.  The numerical action orbit reconstructions assume galactic rotation of 239~\kms, essentially the solution advocated by \citet{2012ApJ...753....8V}.  The radial motions of the void galaxies that we observe depend on our orbit in it's dance with M31.  Here we chose a solution that agrees with the \citet{2012ApJ...753....7S} constraint of minimal transverse motion in the approach of M31 (our model proper motion of M31 is 9 $\mu$arcsec yr$^{-1}$).  With this solution, the Milky Way and M31 are descending in SGZ with velocities $U_Z$ of $-216$ and $-257$~\kms\ respectively, comparable with the median ($-248$~\kms) and mean ($-213$~\kms) of the descent velocities of our neighbors in the Local Sheet within 5 Mpc.

In the Milky Way reference frame, ALFAZOAJ1952+1428 has an observed velocity of 483~\kms\ and a peculiar velocity of $-146$~\kms\ if H$_0 = 75$~\kmsMpc.  The model anticipates a motion of 570~\kms, hence a peculiar velocity of $-59$~\kms.  The observed value is $-87$~\kms\ different from the model value.

This difference between model and observations is of marginal significance.  The error budget is $\pm42$~\kms\ due to the distance uncertainty, $\pm14$~\kms\ due to the uncertainty in H$_0$, and $\pm20$~\kms\ in the averaged negative SGZ motion of nearby galaxies in the Local Sheet.  The joint uncertainty is $\sim \pm50$~\kms.

The situation with KK246 is similar.  The observed velocity is 470~\kms\ in the Milky Way frame, whence a peculiar velocity of $-51$~\kms.  The model predicts 509~\kms\ and a peculiar velocity of $-12$~\kms.  The observed value differs from the model by $-39$~\kms\ with a joint uncertainty of $\sim \pm30$~\kms.

\section{Conclusions}

ALFAZOAJ1952+1428 joins KK246 in a sample of two galaxies with well established distances in the Local Void.  Measured distances are 8.39 and 6.95 Mpc, respectively.  ALFAZOAJ1952+1428 is particularly interesting because it lies near the line of sight toward the void center.  

The HI envelope around KK246 is extensive but otherwise there is nothing particularly unusual with these dwarf galaxies.  These two galaxies with similar integrated magnitudes at $M_B \sim -14$ are the most isolated galaxies known in the local volume.  Both are experiencing moderate ongoing star formation, adding to established older populations.   

The accurate distances permit an evaluation of the deviations of the galaxies from cosmic expansion.  Their orbits are reconstructed by numerical action methods involving a model following 1385 elements within 38 Mpc and embedded in a description of the density field with influences extending to $z \sim 0.05$ obtained with a Wiener Filter analysis.  Both void galaxies are descending in supergalactic SGZ at high velocities, away from the Local Void center, and tending in supergalactic SGY toward the Virgo Cluster.  The negative SGZ motions of  ALFAZOAJ1952+1428 and KK246 respectively are $-146$ and $-51$~\kms\ with respect to the Milky Way.  However, the Milky Way has a substantial negative SGZ motion; $-216$~\kms\ in the specific model under consideration which is comparable to the $\sim -230$~\kms\ descending motion of the ensemble of our neighbors.  Summing the velocity of ALFAZOAJ1952+1428 toward us and our velocity away from the Local Void, the motion of ALFAZOAJ1952+1428 away from the void center is $\sim 370$~\kms.  An empty spherical void in a universe that is topologically flat with 30\% closure density in matter expands at 18~\kmsMpc\ \citep{2008ApJ...676..184T}.  In this approximation, ALFAZOAJ1952+1428 would be $\sim 20$ Mpc $\sim1,500$~\kms\ removed from the void center (placing us $\sim30$~Mpc from the void center).  The direction is obscured by the Milky Way but, while the Local Void is distinctly non-spherical, it does appear to approximate the required extent.
The geometric projection of the motion of KK246 away from the void is less favorable but we measure a motion of $\sim 270$~\kms.  There is the hint that the motions of the void galaxies away from the void center slightly exceed the predictions of our numerical action model but the residuals are not statistically significant.

\bigskip\noindent
This research has been supported by an award from the Space Telescope Science Institute for analysis of observations with Hubble Space Telescope made in the course of program GO-14167. 
IK acknowledges the support of the Russian Science Foundation grant 14-12-00965.

\bibliography{alfazoa}
\bibliographystyle{apj}

\end{document}